\documentclass[twoside,slac_one]{revtex4}
\usepackage{graphicx}
\usepackage{fancyhdr}
\usepackage{amsmath} 
\usepackage{bm}
\usepackage{amsxtra}
\usepackage{amssymb}
\usepackage{amsthm}
\usepackage{latexsym}
\usepackage{lscape}

\def\beq{\begin{equation}}
\def\eeq{\end{equation}}
\def\beqa{\begin{eqnarray}}
\def\eeqa{\end{eqnarray}}

\begin{document}

\title{Second-Order Approximate Corrections for QCD Processes}

%

\author{Nikolaos Kidonakis}
\affiliation{Kennesaw State University, Physics \#1202, Kennesaw, GA 30144, USA}

\begin{abstract}
I present generalized formulas for approximate 
corrections to QCD hard-scattering cross sections through second order in the 
perturbative expansion. The approximate results are based on recent two-loop 
calculations for soft and collinear emission near threshold and are 
illustrated by several applications to strong-interaction processes in 
hadron colliders.
\end{abstract}

\maketitle

\thispagestyle{fancy}


\section{Introduction}

Higher-order perturbative QCD corrections are necessary to produce reliable
estimates of theoretical cross sections and reduce theoretical uncertainties 
for hard-scattering processes.
Collinear and soft-gluon corrections are an important subset of the QCD corrections 
and they can be significant for many processes, particularly near partonic threshold.
The soft and collinear corrections can be derived from factorization theorems 
and resummation formalisms and they require loop calculations 
in the eikonal approximation. In this proceedings I discuss 
resummation, recent calculations of soft anomalous dimensions through 
two loops for a number of processes, and I present master formulas for 
NNLO expansions of the resummed cross section.
Collinear and soft-gluon corrections have been calculated for many 
processes and I present some explicit applications to top quark and 
electroweak boson production. 

In Section 2 the resummation formalism is described and expressions for the resummed 
cross section are presented, including two-loop expressions for some 
universal functions in the resummation. In Section 3 we use the expansion of the 
resummed cross section to derive master formulas for the approximate NLO 
and NNLO corrections.
In section 4 we describe in general the calculation of soft anomalous 
dimensions, and in 
particular the massive cusp anomalous dimension.
In Section 5 we present results for the soft anomalous dimension matrices 
for top-antitop production, in Section 6 for $t$-channel single top quark production, 
in Section 7 for $s$-channel single top quark production, in Section 8 for the 
associated production of a top quark with a $W$ boson (or a charged Higgs), 
and in Section 9 for 
electroweak boson ($W$, $Z$, $\gamma$) production at large transverse momentum.

\section{Higher-order collinear and soft corrections}

Soft-gluon corrections arise from incomplete cancellations of infrared 
divergences between virtual diagrams and real diagrams with soft gluons, 
i.e. low-energy gluons.
These soft corrections take the form 
$[(\ln^k(s_4/M^2))/s_4]_+$ 
where $k \le 2n-1$ for the $\alpha_s^n$ corrections, 
$M$ is a hard scale, and $s_4$ is the kinematical distance 
from partonic threshold.
The leading logarithms are of double collinear and soft emission origin.
There are also purely collinear terms of the form 
$(1/M^2) \ln^k(s_4/M^2)$.

These corrections can be resummed to all orders in perturbation theory.
An essential ingredient in these calculations are the soft anomalous 
dimensions.
At NLL accuracy the resummation requires one-loop  
calculations for the soft anomalous dimensions in the eikonal approximation.
There are several recent results at NNLL with two-loop calculations completed. 

Approximate NNLO cross sections are derived from the expansion of the 
resummed cross section.
These are useful because soft-gluon corrections are dominant near threshold 
and thus the NNLO soft-gluon corrections are expected to approximate 
well the complete NNLO corrections for many processes of interest.  
Also, typically corrections beyond NNLO are small.

Resummation follows from factorization properties of the 
cross section, performed in moment space \cite{NKGS}:
$\sigma=(\prod \psi) \; H_{IL} \, S_{LI} \; (\prod J)$ 
where $\psi$ are distributions for the incoming partons that absorb universal 
collinear singularities, 
$H$ is the hard-scattering matrix ($IL$ are color indices), $S$ is a soft-gluon matrix describing
non-collinear soft-gluon emission, and $J$ are jet functions for the final state. 
We use renormalization group evolution (RGE) to evolve the soft-gluon function 
$$
\left(\mu \frac{\partial}{\partial \mu}
+\beta(g_s) \frac{\partial}{\partial g_s}\right)\,S_{LI}
=-(\Gamma^\dagger_S)_{LB}S_{BI}-S_{LA}(\Gamma_S)_{AI}
$$
where $\beta(g_s)$ is the QCD beta function, with $g_s^2=4\pi\alpha_s$, and  
$\Gamma_S$ is the soft anomalous dimension \cite{NKGS}, a matrix in 
color space and a function of kinematical invariants $s$, $t$, $u$.

The resummed cross section follows from RGE and can be written as 
\beqa
{\hat{\sigma}}^{res}(N) &=&
\exp\left[ \sum_i E_i(N_i)\right] \, \exp\left[ \sum_j E'_j(N')\right]\;
\exp \left[\sum_i 2 \int_{\mu_F}^{\sqrt{s}} \frac{d\mu}{\mu}\;
\gamma_{i/i}\left({\tilde N}_i, \alpha_s(\mu)\right)\right] \;
\nonumber\\ && \hspace{-20mm} \times \,
{\rm tr} \left\{H\left(\alpha_s(\sqrt{s})\right)
\exp \left[\int_{\sqrt{s}}^{{\sqrt{s}}/{\tilde N'}}
\frac{d\mu}{\mu} \;
\Gamma_S^{\dagger}\left(\alpha_s(\mu)\right)\right] \;
S \left(\alpha_s\left(\frac{\sqrt{s}}{\tilde N'}\right)
\right) \;
\exp \left[\int_{\sqrt{s}}^{{\sqrt{s}}/{\tilde N'}}
\frac{d\mu}{\mu}\; \Gamma_S
\left(\alpha_s(\mu)\right)\right] \right\} 
\label{resCS}
\eeqa
with $N$ the Melin moment variable.

The collinear and soft radiation from incoming partons is resummed by
the first exponential in Eq. (\ref{resCS}) with 
\beq
E_i(N_i)=
\int^1_0 dz \frac{z^{N_i-1}-1}{1-z}\;
\left \{\int_1^{(1-z)^2} \frac{d\lambda}{\lambda}
A_i\left(\alpha_s(\lambda s)\right)
+D_i\left[\alpha_s((1-z)^2 s)\right]\right\} \, .
\label{Ei}
\eeq
Purely collinear terms can be derived by replacing $(z^{N-1}-1)/(1-z)\;$ 
in the above equation by $\;- z^{N-1}$.
In Eq. (\ref{Ei}), $A_i$ has the perturbative expansion
$A_i=(\alpha_s/\pi)A_i^{(1)}+(\alpha_s/\pi)^2 A_i^{(2)}+\cdots$, where   
$A_i^{(1)}=C_i$ \cite{GS87} with $C_i=C_F=(N_c^2-1)/(2N_c)$ 
for a quark 
or antiquark and $C_i=C_A=N_c$ for a gluon, with $N_c=3$ the number of colors,
while $A_i^{(2)}=C_i K/2$ \cite{CT89} with 
$K= C_A\; ( 67/18-\pi^2/6 ) - 5n_f/9$,
where $n_f$ is the number of quark flavors.            
Also $D_i=(\alpha_s/\pi)D_i^{(1)}+(\alpha_s/\pi)^2 D_i^{(2)}+\cdots$, 
with $D_i^{(1)}=0$ in Feynman gauge ($D_i^{(1)}=-C_i$ in axial gauge).
In Feynman gauge the two-loop result is (c.f. \cite{CLS97})
\beq
D_i^{(2)}=C_i C_A \left(-\frac{101}{54}+\frac{11}{6} \zeta_2
+\frac{7}{4}\zeta_3\right)
+C_i n_f \left(\frac{7}{27}-\frac{\zeta_2}{3}\right) \, .
\eeq

Collinear and soft radiation from outgoing massless quarks and gluons is resummed by
the second exponential in Eq. (\ref{resCS}) with 
\beq
{E'}_j(N')=
\int^1_0 dz \frac{z^{N'-1}-1}{1-z}\;
\left \{\int^{1-z}_{(1-z)^2} \frac{d\lambda}{\lambda}
A_j \left(\alpha_s\left(\lambda s\right)\right)
+B_j\left[\alpha_s((1-z)s)\right]
+D_j\left[\alpha_s((1-z)^2 s)\right]\right\} \, .
\label{Ej}
\eeq
Here $B_j=(\alpha_s/\pi)B_j^{(1)}+(\alpha_s/\pi)^2 B_j^{(2)}+\cdots$
with $B_q^{(1)}=-3C_F/4$ and $B_g^{(1)}=-\beta_0/4$ 
\cite{GS87,CT89}, where
$\beta_0$ is the lowest-order $\beta$-function, $\beta_0=(11C_A-2n_f)/3$.
Also  
\beq
B_q^{(2)}=C_F^2\left(-\frac{3}{32}+\frac{3}{4}\zeta_2-\frac{3}{2}\zeta_3\right)
+C_F C_A \left(-\frac{1539}{864}-\frac{11}{12}\zeta_2+\frac{3}{4}\zeta_3\right)
+n_f C_F \left(\frac{135}{432}+\frac{\zeta_2}{6}\right)
\eeq
\beq
B_g^{(2)}=C_A^2\left(-\frac{1025}{432}-\frac{3}{4}\zeta_3\right)
+\frac{79}{108} C_A \, n_f +C_F \frac{n_f}{8}-\frac{5}{108} n_f^2 \, .
\eeq

The factorization scale, $\mu_F$, dependence is controlled by
the moment-space anomalous dimension 
$\gamma_{i/i}=-A_i \ln {\tilde N}_i +\gamma_i$ with $\gamma_i$ the parton 
anomalous dimensions \cite{GALY79,GFP80}.
We have 
\beq
\gamma_i=(\alpha_s/\pi) \gamma_i^{(1)}
+(\alpha_s/\pi)^2 \gamma_i^{(2)} + \cdots
\eeq
with  $\gamma_q^{(1)}=3C_F/4$, $\gamma_g^{(1)}=\beta_0/4$,
\beq
\gamma_q^{(2)}=C_F^2\left(\frac{3}{32}-\frac{3}{4}\zeta_2
+\frac{3}{2}\zeta_3\right)
+C_F C_A\left(\frac{17}{96}+\frac{11}{12}\zeta_2-\frac{3}{4}\zeta_3\right)
+n_f C_F \left(-\frac{1}{48}-\frac{\zeta_2}{6}\right)\, ,
\eeq
and
\beq
\gamma_g^{(2)}=C_A^2\left(\frac{2}{3}+\frac{3}{4}\zeta_3\right)
-n_f\left(\frac{C_F}{8}+\frac{C_A}{6}\right) \, .
\eeq

The noncollinear soft gluon emission is controlled by the 
soft anomalous dimension $\Gamma_S$.  
We determine $\Gamma_S$ from the coefficients of ultraviolet poles in 
dimensionally regularized eikonal diagrams \cite{NKGS,NK2l}.
The hard $H$ and soft $S$ matrices as well as 
$\Gamma_S$ are process-dependent and thus have to be calculated separately for 
each hard-scattering process.

\section{NNLO approximate cross sections}

The resummed cross section in Eq. (\ref{resCS}) can be expanded to any finite order. Here we present 
master fromulas for the NLO and NNLO expansions.

We define ${\cal D}_k(s_4) \equiv \left[(\ln^k(s_4/M^2))/s_4\right]_+$. 
Then the NLO approximate corrections are
\beq
{\hat{\sigma}}^{(1)} = \sigma^B \frac{\alpha_s(\mu_R)}{\pi}
\left\{c_3\, {\cal D}_1(s_4) + c_2\,  {\cal D}_0(s_4) 
+c_1\,  \delta(s_4)\right\}+\frac{\alpha_s^{d_{\alpha_s}+1}(\mu_R^2)}{\pi} 
\left[A^c \, {\cal D}_0(s_4)+T_1^c \, \delta(s_4)\right] 
\eeq
where $\sigma^B$ is the Born term, $\mu_R$ is the renormalization scale, 
and $d_{\alpha_s}$ denotes the power of $\alpha_s$ for the leading-order 
cross section. Also 
$c_3=\sum_i 2 \, A_i^{(1)} -\sum_j A_j^{(1)}$ 
and $c_2=c_2^{\mu}+T_2$, 
with
$c_2^{\mu}=-\sum_i A_i^{(1)} \ln\left(\mu_F^2/M^2\right)$
and
\beq
T_2=\sum_i \left[
-2 \, A_i^{(1)} \, \ln\left(\frac{-t_i}{M^2}\right)+D_i^{(1)}
-A_i^{(1)} \ln\left(\frac{M^2}{s}\right)\right]
+\sum_j \left[B_j^{(1)}+D_j^{(1)}
-A_j^{(1)} \, \ln\left(\frac{M^2}{s}\right)\right] \, .
\eeq
Also
\beq
A^c={\rm tr} \left(H^{(0)} \Gamma_S^{(1)\,\dagger} S^{(0)}
+H^{(0)} S^{(0)} \Gamma_S^{(1)}\right) \, .
\eeq

We split the $c_1$ coefficient as 
$c_1 =c_1^{\mu} +T_1$, with
\beq
c_1^{\mu}=\sum_i \left[A_i^{(1)}\, \ln\left(\frac{-t_i}{M^2}\right) 
-\gamma_i^{(1)}\right]\ln\left(\frac{\mu_F^2}{M^2}\right)
+d_{\alpha_s} \frac{\beta_0}{4} \ln\left(\frac{\mu_R^2}{M^2}\right) 
\eeq
the terms involving the factorization scale $\mu_F$ and 
renormalization scale $\mu_R$ dependence.
On the other hand $T_1$ as well as $T_1^c$ are not derivable from 
resummation but can be extracted from complete NLO calculations.

The NNLO approximate corrections are
\beqa
{\hat{\sigma}}^{(2)}&=&\sigma^B \frac{\alpha_s^2(\mu_R)}{\pi^2}
\left\{\frac{1}{2}c_3^2\, {\cal D}_3(s_4) + 
\left[\frac{3}{2}c_3 c_2-\frac{\beta_0}{4} c_3
+\sum_j \frac{\beta_0}{8} A_j^{(1)}\right]  {\cal D}_2(s_4) \right.
\nonumber \\ &&  
{}+\left[c_3 c_1+c_2^2-\zeta_2 c_3^2
-\frac{\beta_0}{2} T_2+\frac{\beta_0}{4} c_3 
\ln\left(\frac{\mu_R^2}{M^2}\right)+\sum_i 2A_i^{(2)}-\sum_j A_j^{(2)}
+\sum_j \frac{\beta_0}{4} B_j^{(1)}\right] {\cal D}_1(s_4)
\nonumber \\ &&  
{}+\left[c_2 c_1-\zeta_2 c_3 c_2+\zeta_3 c_3^2
+\frac{\beta_0}{4} c_2 \ln\left(\frac{\mu_R^2}{s}\right) \right. 
-\sum_i \frac{\beta_0}{2} A_i^{(1)} \ln^2\left(\frac{-t_i}{M^2}\right)
\nonumber \\ &&  
{}+\sum_i [\left(-2 A_i^{(2)}+\frac{\beta_0}{2} D_i^{(1)}\right) 
\ln\left(\frac{-t_i}{M^2}\right)+D_i^{(2)}
+\frac{\beta_0}{8}  A_i^{(1)} \ln^2\left(\frac{\mu_F^2}{s}\right) 
-A_i^{(2)} \ln\left(\frac{\mu_F^2}{s}\right)]
\nonumber \\ &&  \left. \left.
{}+\sum_j [B_j^{(2)}+D_j^{(2)}-\left(A_j^{(2)}+\frac{\beta_0}{4}
(B_j^{(1)}+2 D_j^{(1)})\right) \ln\left(\frac{M^2}{s}\right)
+\frac{3\beta_0}{8} A_j^{(1)} \ln^2\left(\frac{M^2}{s}\right)]\right] 
 {\cal D}_0(s_4) \right\}
\nonumber \\ && 
{}+\frac{\alpha_s^{d_{\alpha_s}+2}(\mu_R)}{\pi^2} 
\left\{\frac{3}{2} c_3 A^c \, {\cal D}_2(s_4)
+\left[\left(2 c_2-\frac{\beta_0}{2}\right) A^c+c_3 T_1^c +F^c\right]
{\cal D}_1(s_4) \right. 
\nonumber \\ && \left. 
{}+\left[\left(c_1-\zeta_2 c_3+\frac{\beta_0}{4}\ln\left(\frac{\mu_R^2}{s}
\right)\right)A^c+c_2 T_1^c +F^c \ln\left(\frac{M^2}{s}\right) +G^c\right]
{\cal D}_0(s_4) \right\} 
\label{NNLO}
\eeqa
where
\beq
F^c={\rm tr} \left[H^{(0)} \left(\Gamma_S^{(1)\,\dagger}\right)^2 S^{(0)}
+H^{(0)} S^{(0)} \left(\Gamma_S^{(1)}\right)^2
+2 H^{(0)} \Gamma_S^{(1)\,\dagger} S^{(0)} \Gamma_S^{(1)} \right] 
\eeq
and
\beqa
G^c&=&{\rm tr} \left[H^{(1)} \Gamma_S^{(1)\,\dagger} S^{(0)}
+H^{(1)} S^{(0)} \Gamma_S^{(1)} + H^{(0)} \Gamma_S^{(1)\,\dagger} S^{(1)}
+H^{(0)} S^{(1)} \Gamma_S^{(1)} \right.
\nonumber \\ && \quad \quad \left.
{}+H^{(0)} \Gamma_S^{(2)\,\dagger} S^{(0)}
+H^{(0)} S^{(0)} \Gamma_S^{(2)} \right] \, .
\eeqa

In Eq. (\ref{NNLO}), $c_3$, $c_2$, $c_1$, etc are defined from the NLO 
expansion.
The two-loop universal quantities $A^{(2)}$, $B^{(2)}$, $D^{(2)}$ are known and were 
presented in the previous section.
The functions $H$, $S$, $\Gamma_S$ are process dependent and have been 
calculated for many processes.

\section{Soft anomalous dimensions}

The two-loop process-dependent soft anomalous dimensions, $\Gamma_S^{(2)}$,
have been recently calculated for several processes.
The calculations use the eikonal approximation 
where the rules for soft gluon emission simplify. For the emission of a gluon 
with momentum $k$ off a quark with momentum $p$ the Feynman rules simplify 
as follows in the limit $k \rightarrow 0$: 
$${\bar u}(p) \, (-i g_s T_F^c) \, \gamma^{\mu}
\frac{i (p\!\!/+k\!\!/+m)}{(p+k)^2
-m^2+i\epsilon} \rightarrow {\bar u}(p)\,  g_s T_F^c \, \gamma^{\mu}
\frac{p\!\!/+m}{2p\cdot k+i\epsilon}
={\bar u}(p)\, g_s T_F^c \,
\frac{v^{\mu}}{v\cdot k+i\epsilon} $$
with $v$ a dimensionless vector proportional to the momentum $p$, and 
$T_F^c$ the generators of $SU(3)$ in the fundamental representation.

We perform the calculations of eikonal diagrams in momentum space 
and Feynman gauge.
Complete two-loop results for the soft (cusp) anomalous dimension for 
$e^+ e^- \rightarrow t {\bar t}$ were presented in \cite{NK2l}.

\begin{figure}[h]
\centering
\includegraphics[width=80mm]{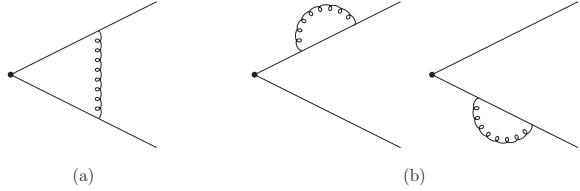}
\caption{One-loop cusp diagrams.}
\label{loopg1}
\end{figure}

In Fig. \ref{loopg1} we show the one-loop diagrams for the massive 
cusp anomalous dimension, with the lines representing a top and antitop
pair in the process $e^+ e^- \rightarrow t {\bar t}$. From the UV poles
of these diagrams we find the one-loop result 
\beq
\Gamma_S^{(1)}=C_F \left[-\frac{(1+\beta^2)}{2\beta} 
\ln\left(\frac{1-\beta}{1+\beta}\right) -1\right]
\label{Gamma1}
\eeq
where $\beta=\sqrt{1-4m^2/s}$.

\begin{figure}[h]
\centering
\includegraphics[width=80mm]{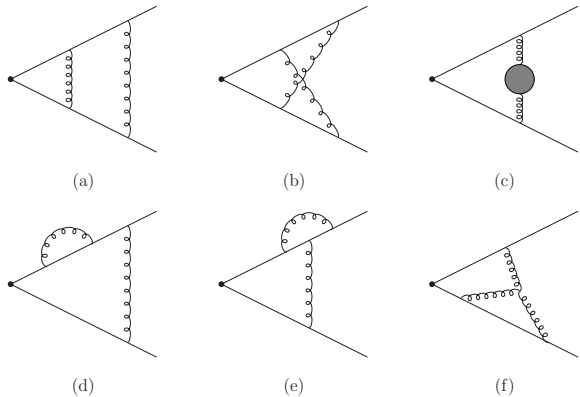}
\caption{Two-loop cusp diagrams.}
\label{loopg2}
\end{figure}

In Fig. \ref{loopg2} we show two-loop cusp diagrams (there are additional 
ones involving top-quark self energies). From the UV poles
of these diagrams we find the two-loop result \cite{NK2l} 
\beqa
&& \hspace{-5mm}\Gamma_S^{(2)}=\frac{K}{2} \, \Gamma_S^{(1)}
+C_F C_A \left\{\frac{1}{2}+\frac{\zeta_2}{2}
+\frac{1}{2} \ln^2\left(\frac{1-\beta}{1+\beta}\right) \right.
\nonumber \\ && \hspace{5mm}
{}-\frac{(1+\beta^2)^2}{8 \beta^2} \left[\zeta_3
+\zeta_2 \ln\left(\frac{1-\beta}{1+\beta}\right)
+\frac{1}{3} \ln^3\left(\frac{1-\beta}{1+\beta}\right) 
+\ln\left(\frac{1-\beta}{1+\beta}\right) 
{\rm Li}_2\left(\frac{(1-\beta)^2}{(1+\beta)^2}\right) 
-{\rm Li}_3\left(\frac{(1-\beta)^2}{(1+\beta)^2}\right)\right] 
\nonumber \\ &&  \hspace{5mm} 
{}-\frac{(1+\beta^2)}{4 \beta} \left[\zeta_2
-\zeta_2 \ln\left(\frac{1-\beta}{1+\beta}\right) 
+\ln^2\left(\frac{1-\beta}{1+\beta}\right)
-\frac{1}{3} \ln^3\left(\frac{1-\beta}{1+\beta}\right)
+2  \ln\left(\frac{1-\beta}{1+\beta}\right) 
\ln\left(\frac{(1+\beta)^2}{4 \beta}\right) \right.
\nonumber \\ &&  \hspace{25mm} \left. \left. 
{}-{\rm Li}_2\left(\frac{(1-\beta)^2}{(1+\beta)^2}\right)\right]\right\} \, .
\label{Gamma2}
\eeqa

The results in Eqs. (\ref{Gamma1}) and (\ref{Gamma2}) can also be expressed 
in terms of the cusp angle. Such expressions have been presented 
in \cite{NK2l} and are more analytically explicit than earlier 
results in \cite{KorRad}. 

More recently results have appeared for $t{\bar t}$ hadroproduction, 
$t$-channel single top production,  $s$-channel single top production, 
$bg \rightarrow t W^-$ and $bg \rightarrow t H^-$, and
direct photon, $W$, and $Z$ production at large $p_T$.
The color structure gets more complicated with more than two colored partons 
in the process. The cusp anomalous dimension is an essential component of 
other calculations.

\section{Top-antitop production in hadron colliders}

The soft anomalous dimension matrix for $q{\bar q} \rightarrow t{\bar t}$ is
a $2\times 2$ matrix: 
\beqa
\Gamma_{S\, q{\bar q}}=\left[\begin{array}{cc}
\Gamma_{q{\bar q} \, 11} & \Gamma_{q{\bar q} \, 12} \\
\Gamma_{q{\bar q} \, 21} & \Gamma_{q{\bar q} \, 22}
\end{array}
\right] \, .
\nonumber
\eeqa
Results have been presented in \cite{NKtop} in a singlet-octet color exchange basis through two loops.
At one loop
\beqa
&& \Gamma_{q{\bar q} \,11}^{(1)}=-C_F \, [L_{\beta}+1] 
\hspace{15mm}
\Gamma_{q{\bar q} \,21}^{(1)}=
2\ln\left(\frac{u_1}{t_1}\right) \hspace{15mm}
\Gamma_{q{\bar q} \,12}^{(1)}=
\frac{C_F}{C_A} \ln\left(\frac{u_1}{t_1}\right) 
\nonumber \\ &&
\Gamma_{q{\bar q} \,22}^{(1)}=C_F
\left[4\ln\left(\frac{u_1}{t_1}\right)
-L_{\beta}-1\right]
+\frac{C_A}{2}\left[-3\ln\left(\frac{u_1}{t_1}\right)
+\ln\left(\frac{t_1u_1}{s m^2}\right)+L_{\beta}\right]
\nonumber
\eeqa
where 
$L_{\beta}=\frac{1+\beta^2}{2\beta}\ln\left(\frac{1-\beta}{1+\beta}\right)$
with $\beta=\sqrt{1-4m^2/s}$.

We write the two-loop cusp anomalous dimension, Eq. (\ref{Gamma2}), 
from the previous section as
$\Gamma_S^{(2)}=\frac{K}{2} \, \Gamma_S^{(1)}+C_F C_A M_{\beta}$ where 
$M_{\beta}$ denotes the terms in curly brackets in Eq. (\ref{Gamma2}).
We use $M_{\beta}$ below in the two-loop matrix for $q{\bar q} 
\rightarrow t{\bar t}$. 
Then at two loops for $q{\bar q} \rightarrow t{\bar t}$ we find \cite{NKtop}
\beqa
&& \Gamma_{q{\bar q} \,11}^{(2)}=\frac{K}{2} \Gamma_{q{\bar q} \,11}^{(1)}
+C_F C_A \, M_{\beta} \hspace{20mm}
\Gamma_{q{\bar q} \,22}^{(2)}=
\frac{K}{2} \Gamma_{q{\bar q} \,22}^{(1)}
+C_A\left(C_F-\frac{C_A}{2}\right) \, M_{\beta} 
\nonumber \\ &&
\Gamma_{q{\bar q} \,21}^{(2)}=
\frac{K}{2}  \Gamma_{q{\bar q} \,21}^{(1)} +C_A N_{\beta} \ln\left(\frac{u_1}{t_1}\right) \hspace{15mm} 
\Gamma_{q{\bar q} \,12}^{(2)}=
\frac{K}{2} \Gamma_{q{\bar q} \,12}^{(1)} -\frac{C_F}{2} N_{\beta} \ln\left(\frac{u_1}{t_1}\right) 
\nonumber
\eeqa
with $N_{\beta}$ a subset of terms of $M_{\beta}$,
\beq
N_{\beta}=\frac{1}{2} \ln^2\left(\frac{1-\beta}{1+\beta}\right)
-\frac{(1+\beta^2)}{4 \beta} \left[
\ln^2\left(\frac{1-\beta}{1+\beta}\right)
+2  \ln\left(\frac{1-\beta}{1+\beta}\right) 
\ln\left(\frac{(1+\beta)^2}{4 \beta}\right) 
-{\rm Li}_2\left(\frac{(1-\beta)^2}{(1+\beta)^2}\right)\right].
\eeq

Corresponding results for the $gg\rightarrow t{\bar t}$ channel 
are given in \cite{NKtop}.

\section{Single top quark production - $t$ channel}

The dominant single top production channel at both Tevatron and LHC energies 
is the $t$ channel, which at lowest order involves the partonic processes
$qb \rightarrow q' t$ and ${\bar q}b \rightarrow {\bar q}' t$.  
Here we show some of the elements of the one-loop  
and two-loop \cite{NKtch}
soft anomalous dimension matrices that are used in the NNLO expansions.
At one loop 
$$
{\Gamma}_{S\, 11}^{(1)}
=C_F \left[\ln\left(\frac{-t}{s}\right)
+\ln\left(\frac{m_t^2-t}{m_t\sqrt{s}}\right)-\frac{1}{2}\right] 
$$
$$
\Gamma_{S\, 21}^{(1)}=\ln\left(\frac{u(u-m_t^2)}{s(s-m_t^2)}\right)
\hspace{2cm}
\Gamma_{S\, 12}^{(1)}=\frac{C_F}{2N_c} \, \Gamma_{S\,21}^{(1)} \, .
$$
At two loops we find 
$$
\Gamma_{S\, 11}^{(2)}=\frac{K}{2}\Gamma_{S\, 11}^{(1)}
+C_F C_A \frac{(1-\zeta_3)}{4} \, .
$$

\section{Single top quark production - $s$ channel}

\begin{figure}[h]
\centering
\includegraphics[width=80mm]{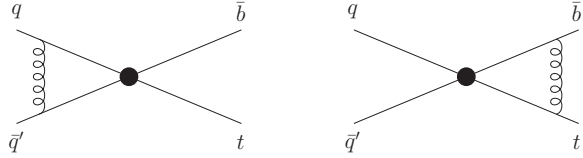}
\caption{One-loop $s$-channel diagrams.}
\label{s1loop}
\end{figure}

The $s$-channel processes are of the form $q{\bar q}' \rightarrow {\bar b} t$.
One-loop eikonal diagrams for $s$-channel production are shown 
in Fig. \ref{s1loop}.

\begin{figure*}[t]
\centering
\includegraphics[width=50mm]{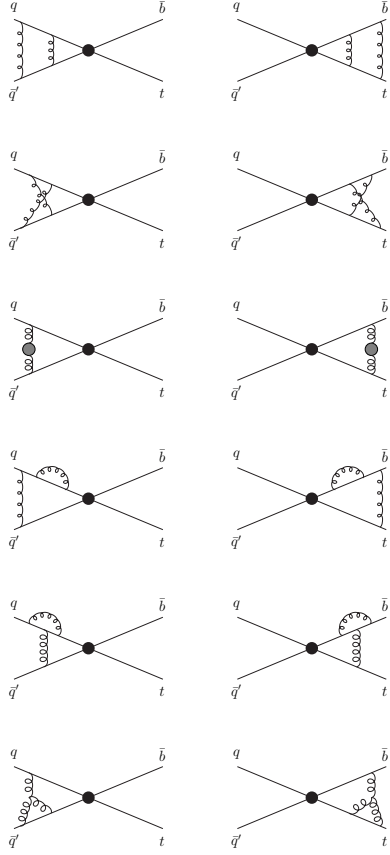}
\caption{Two-loop $s$-channel diagrams.}
\label{s2loopv}
\end{figure*} 

\begin{figure*}[t]
\centering
\includegraphics[width=50mm]{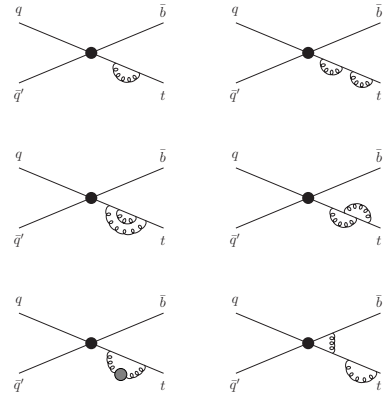}
\caption{One- and two-loop top-quark self-energy graphs in the $s$ channel.}
\label{s2loopse}
\end{figure*} 

Two-loop eikonal diagrams are shown in Figs. \ref{s2loopv} and 
\ref{s2loopse}.

The 11 elements of the soft anomalous dimension for $s$-channel single top 
production at one and two loops are \cite{NKsch}
$$
\Gamma_{S\, 11}^{(1)}=C_F \left[\ln\left(\frac{s-m_t^2}{m_t\sqrt{s}}\right)
-\frac{1}{2}\right]\, , \quad \quad
\Gamma_{S\, 11}^{(2)}=\frac{K}{2} \Gamma_{S\, 11}^{(1)}
+C_F C_A \frac{(1-\zeta_3)}{4} \, .
$$

\section{Associated production of a top quark with a $W^-$ or $H^-$}

The associated production of a top quark with a $W$ boson has the second 
largest cross section among single-top quark processes at the LHC. 
The NNLL resummation for this process was derived in \cite{NKtW}.
The two-loop eikonal diagrams that are evaluated in the calculation of the 
two-loop soft anomalous dimension are shown in Ref. \cite{NKtW}. 

The soft anomalous dimension for $bg \rightarrow tW^-$ 
(or $bg \rightarrow tH^-$) is at one loop 
$$
\Gamma_{S,\, tW^-}^{(1)}=C_F \left[\ln\left(\frac{m_t^2-t}{m_t\sqrt{s}}\right)
-\frac{1}{2}\right] +\frac{C_A}{2} \ln\left(\frac{m_t^2-u}{m_t^2-t}\right)
$$
and at two loops 
$$
\Gamma_{S,\, tW^-}^{(2)}=\frac{K}{2} \Gamma_{S,\, tW^-}^{(1)}
+C_F C_A \frac{(1-\zeta_3)}{4} \, .
$$

\section{$W$, $Z$, and direct photon production at large $p_T$}

Threshold corrections for electroweak boson production 
dominate the $p_T$ distribution at large transverse 
momentum. NLL resummation for $W$ production was studied 
in \cite{NKVD,NKASV}. 

The two loop soft anomalous dimensions for NNLL resummation can be derived 
from two-loop diagrams in the eikonal approximation \cite{NKren}.
For $qg\rightarrow Wq$ (or $qg\rightarrow Zq$ or $qg\rightarrow \gamma q$)
the diagrams are similar to those for $tW$ production. The results for 
the soft anomalous dimension at one and two loops \cite{NKren} are
$$
\Gamma_{S,\, qg\rightarrow Wq}^{(1)}=C_F \ln\left(\frac{-u}{s}\right)
+\frac{C_A}{2} \ln\left(\frac{t}{u}\right)
$$
and 
$$
\Gamma_{S,\, qg \rightarrow Wq}^{(2)}=\frac{K}{2} \Gamma_{S,\, qg \rightarrow Wq}^{(1)} \, .
$$

For $q {\bar q}\rightarrow Wg$ (or $q {\bar q}\rightarrow Zg$
or $q {\bar q}\rightarrow \gamma g$) the corresponding one-loop and two-loop 
expressions are \cite{NKren} 
$$
\Gamma_{S,\, q{\bar q}\rightarrow Wg}^{(1)}=\frac{C_A}{2} \ln\left(\frac{tu}{s^2}\right)$$
and
$$
\Gamma_{S,\, q{\bar q} \rightarrow Wg}^{(2)}=\frac{K}{2} \Gamma_{S,\, q{\bar q} \rightarrow Wg}^{(1)} \, .
$$

\begin{acknowledgments}
This work was supported by the National Science Foundation under 
Grant No. PHY 0855421.
\end{acknowledgments}

\bigskip 

\begin{thebibliography}{99} 

\bibitem{NKGS}
N. Kidonakis and G. Sterman, Nucl. Phys. B {\bf 505}, 321 (1997) [hep-ph/9705234]. 

\bibitem{GS87}
G. Sterman, Nucl. Phys. B {\bf 281}, 310 (1987).

\bibitem{CT89}
S. Catani and L. Trentadue, Nucl. Phys. B {\bf 327}, 323 (1989).

\bibitem{CLS97}
H. Contopanagos, E. Laenen, and G. Sterman, Nucl. Phys. B {\bf 484}, 303 (1997)
[hep-ph/9604313].

\bibitem{GALY79}
A. Gonzalez-Arroyo, C. Lopez, and F.J. Yndurain,
Nucl. Phys. B {\bf 153}, 161 (1979).
 
\bibitem{GFP80}
G. Curci, W. Furmanski, and R. Petronzio, Nucl. Phys. B {\bf 175}, 27 (1980).

\bibitem{NK2l}
N. Kidonakis, Phys. Rev. Lett. {\bf 102}, 232003 (2009)
[arXiv:0903.2561 [hep-ph]].

\bibitem{KorRad}
G.P. Korchemsky and A.V. Radyushkin, Phys. Lett. B {\bf 171}, 459 (1986); 
Nucl. Phys. B {\bf 283}, 342 (1987); Phys. Lett. B {\bf 279}, 359 (1992) 
[hep-ph/9203222].

\bibitem{NKtop}
N. Kidonakis, Phys. Rev. D {\bf 82}, 114030 (2010)
[arXiv:1009.4935 [hep-ph]].

\bibitem{NKtch} 
N. Kidonakis, Phys. Rev. D {\bf 83}, 091503 (2011)
[arXiv:1103.2792 [hep-ph]].

\bibitem{NKsch}
N. Kidonakis, Phys. Rev. D {\bf 81}, 054028 (2010)
[arXiv:1001.5034 [hep-ph]].

\bibitem{NKtW}
N. Kidonakis, Phys.Rev. D {\bf 82}, 054018 (2010)
[arXiv:1005.4451 [hep-ph]].

\bibitem{NKVD}
N. Kidonakis and V. Del Duca, Phys. Lett. B {\bf 480}, 87 (2000) 
[hep-ph/9911460].

\bibitem{NKASV}
N. Kidonakis and A. Sabio Vera,  JHEP {\bf 02}, 027 (2004) [hep-ph/0311266].

\bibitem{NKren}
N. Kidonakis, in {\sl DIS2011}, arXiv:1105.4267 [hep-ph].

\end{thebibliography}

\end{document}